\documentclass[aps,prl,floatfix,superscriptaddress,preprint,showpacs]{revtex4}
\usepackage{times}
\usepackage{dcolumn}                 
\usepackage{graphicx}
\usepackage{epsf}
\usepackage{amsmath}
\usepackage{bm}
\newcolumntype{.}{D{x}{}{-1}}

\newcommand*{\centt}[1]{\multicolumn{1}{c}{#1}}
\newcolumntype{w}[1]{D{.}{.}{#1}}

\begin{document}
\preprint{Version 3.0}

\title{Nuclear polarizability of helium isotopes in atomic transitions}

\author{K. Pachucki}
\affiliation{
Institute of Theoretical Physics, Warsaw University,
Ho\.{z}a 69, 00-681 Warsaw, Poland} 

\author{A.M. Moro}
\affiliation{Departamento de FAMN, Universidad de Sevilla, Apartado 1065,
             41080 Sevilla, Spain }

\begin{abstract}
We estimate the nuclear polarizability correction
to atomic transition frequencies in various helium isotopes. 
This effect is non-negligible for high precision tests of quantum electrodynamics
or accurate determination of the nuclear charge radius from spectroscopic 
measurements in helium atoms and ions, in particular
it amounts to $28(3)$ kHz for 1S-2S transition in $^4$He$^+$.
\end{abstract}

\maketitle
There are various corrections 
which need to be included in an accurate calculation of the 
atomic energy levels. These are relativistic and QED effects, finite nuclear mass
and, to some extent, finite nuclear size. There is also
an additional correction which comes from possible nuclear excitation
due to the electric field of the surrounding electrons. This effect
is usually neglected, as it is relatively small compared to the uncertainties
in the nuclear charge radii. There are however exceptions, where the nuclear
polarizability correction can be significant. The first known example
was the $2P_{3/2}-2S_{1/2}$ transition in muonic helium $\mu$-$^4$He$^+$
\cite{bj,friar_m}, where the polarizability
correction is about 1\% of the finite nuclear size effect. Another example is the
isotope shift in the $1S-2S$ transition frequency between hydrogen and deuterium.
The nuclear polarizability correction of about $20$ kHz, was two orders of
magnitude larger than the experimental precision \cite{huber}, and helped to resolve
experimental discrepancies for the deuteron charge radius. Another
very recent example is the isotope shift in the $2S-3S$ transition frequency 
between $^{11}$Li and $^7$Li \cite{11li}. The  $^{11}$Li nucleus has probably the largest
nuclear polarizability among all light nuclei, with a contribution to the $2S-3S$ isotope
shift of about $36$~kHz. In this paper we study in detail the nuclear
polarizability correction to atomic transitions for helium isotopes:
$^3$He, $^4$He and $^6$He and compare with currently available 
and planned accurate measurements of transition frequencies
in helium atoms and ions. 
  
The interaction of the  nucleus with an electromagnetic field
can be described by the following Hamiltonian:
\begin{equation}
H_{\rm int} = q\,A^0 - \vec d\cdot \vec E  -\vec\mu\cdot\vec B 
-\frac{q}{6}\,\langle r^2\rangle \,\nabla\cdot \vec E  , \label{01}
\end{equation}
which is valid as long as the characteristic momentum of the electromagnetic field
is smaller than the inverse of the nuclear size. Otherwise, one has to use
a complete description in terms of formfactors and structure functions.
Under this assumption, the dominant term for 
the nuclear excitation is the electric dipole interaction. This is the main approximation of 
our approach,  which may not always be valid.
It was checked however that higher order polarizabilities are quite small 
(below 1~kHz) for deuterium \cite{ros,friar}, and this should be similar for He
isotopes. Within this low electromagnetic momentum approximation, the nuclear 
polarizability correction to the energy
is given by the following formula \cite{mp} (in units $\hbar = c = 1, e^2 = 4\,\pi\,\alpha$): 
\begin{equation}
E_{\rm pol} = -m\,\alpha^4\,\Bigl\langle\sum_a\delta^3(r_a)\Bigr\rangle\;
(m^3\,\tilde\alpha_{\rm pol}), \label{02}
\end{equation}
where $m$ is the electron mass and the expectation value of the Dirac $\delta$ 
is taken with the electron wave function in atomic units. 
For hydrogenic systems it is equal to $Z^3/\pi$. In the equation 
above,  $\tilde\alpha_{\rm pol}$  is a {\em weighted} electric polarizability of the nucleus,
which is given by the following double integral
\begin{eqnarray}
\tilde\alpha_{\rm pol} &=& \frac{16\,\alpha}{3}\,\int_{E_T}^\infty dE\,
\frac{1}{e^2}\,|\langle\phi_N|\vec d|E\rangle|^2\,\int_0^\infty\,\frac{d w}{w}\,
\frac{E}{E^2+w^2}
\nonumber \\ &&\hspace*{-5ex}
\times\frac{1}{(\kappa+\kappa^\star)}\,\biggl[1+
\frac{1}{(\kappa+1)(\kappa^\star+1)}\,\biggl(\frac{1}{\kappa+1}+\frac{1}{\kappa^\star+1}
\biggr)\biggr]\label{03}
\end{eqnarray}
where $ \kappa = \sqrt{1+2\,i\,m/w}$ and 
$E_T$ is the excitation energy for the breakup threshold. 
The kets $|\phi_N\rangle$ and $| E \rangle$ denote
the ground state of the nucleus and a dipole excited state with excitation energy $E$, respectively.
The square of the dipole moment is related to the so called $B(E1)$ function by
\begin{equation}
|\langle\phi_N|\vec d|E\rangle|^2 = \frac{4\,\pi}{3}\,\frac{dB(E1)}{dE} \label{04}
\end{equation}
in units $e^2$ fm$^2$ MeV$^{-1}$, which explains the presence of $e^2$ in the
denominator in Eq.~(\ref{03}).  The $B(E1)$ function is, in turn, 
directly related to the 
photoabsorption cross section at photon energies $E$
\begin{equation}
\sigma(E) = \frac{16\,\pi^3}{9}\,\alpha\,E\,
\left(\frac{1}{e^2}\,\frac{d B(E1)}{dE}\right) \label{05}
\end{equation}
which allows us to obtain the $B(E1)$ function from experimental data.

If $E_T$ is much larger than the electron mass $m$, one can perform a
small electron mass expansion and obtain a simplified formula, 
in agreement with that previously derived in \cite{weitz}:
\begin{eqnarray}
\tilde \alpha_{\rm pol} &=& \frac{19}{6}\,\alpha_{\rm E} + 5\,\alpha_{\rm Elog} \label{06}\\
\alpha_{\rm E} &=& \frac{2\,\alpha}{3}\,\frac{1}{e^2}\,\biggl\langle\phi_N\biggl|
\,\vec d\,\frac{1}{H_N-E_N}\,\vec d\;\biggr|\phi_N\biggr\rangle \nonumber \\
&=&\frac{8\,\pi\,\alpha}{9}\,\int_{E_T} \frac{dE}{E}  \,\frac{1}{e^2}\frac{d B(E1)}{dE}\\
\alpha_{\rm Elog} &=& \frac{2\,\alpha}{3}\,\frac{1}{e^2}\,\biggl\langle\phi_N\biggl|
\,\vec d\,\frac{1}{H_N-E_N}\,\ln\biggl(\frac{2(H_N-E_N)}{m}\biggr)\,\vec d
\;\biggr|\phi_N\biggr\rangle \nonumber \\
&=&\frac{8\,\pi\,\alpha}{9}\,\int_{E_T} \frac{dE}{E}  \,\frac{1}{e^2}\frac{d
  B(E1)}{dE}\,\ln\biggl(\frac{2\,E}{m}\biggr)\,,
\end{eqnarray}
where $\alpha_{\rm E}$ is the static electric dipole polarizability and
$\alpha_{\rm Elog}$ is the logarithmically modified polarizability.
We have tested this approximation for $^3$He  and $^4$He isotopes,  and found  that
numerical results differ from the exact formula in Eq.~(\ref{03}) by less than $0.1\%$.

In the opposite situation, i.e.\  when $m$ is much larger that $E_T$, 
that corresponds to the nonrelativistic limit, the
polarizability correction adopts the form (with $m$ being the reduced mass here):
\begin{equation}
E_{\rm pol} = -m\,\alpha^4\,\Bigl\langle\sum_a\delta^3(r_a)\Bigr\rangle\;
\frac{32\,\pi\,\alpha\,m^2}{9}\,\int_{E_T} dE\,\frac{1}{e^2}\frac{d B(E1)}{dE}\,
\sqrt{\frac{m}{2\,E}} . \label{07}
\end{equation}
This approximation is justified for muonic helium atoms or ions, because the muon mass
($\sim106$ MeV) is much larger than the threshold energy $E_T$ [see Eq.~(\ref{08})].
This formula requires, however few significant corrections, namely Coulomb
distortion and formfactor corrections. They were obtained by Friar in \cite{friar_m} for 
the calculation of the polarizability correction in $\mu\,^4$He. 
This nonrelativistic approximation, however, is not valid for electronic atoms
since the typical nuclear excitation energy in light nuclei is larger than
the electron mass. 

We consider in this work three helium isotopes, namely $^3$He, $^4$He and $^6$He, which 
are stable or
sufficiently long lived for performing precise atomic measurements. The separation energy
$S\equiv E_T$ for these helium isotopes are \cite{audi}: 
\begin{eqnarray}
S_{2n}(^6{\rm He}) &=& 0.975\,{\rm MeV}\,, \nonumber \\
S_p(^4{\rm He})    &=& 19.8\,{\rm MeV}\,, \nonumber\\
S_n(^4{\rm He})    &=& 20.6\,{\rm MeV}\,, \label{08} \\
S_p(^3{\rm He})    &=& 5.493\,49\,{\rm MeV}\,, \nonumber\\
S_n(^3{\rm He})&=& 7.718\,04\,{\rm MeV}\,.\nonumber
\end{eqnarray}
The separation energy $S$ is the main factor which determines
the magnitude of the nuclear polarizability correction, since
$E_{\rm pol}$ is approximately proportional to the inverse of $S$.

\begin{figure}[htb]%
\begin{center}\includegraphics[width=0.7\linewidth]{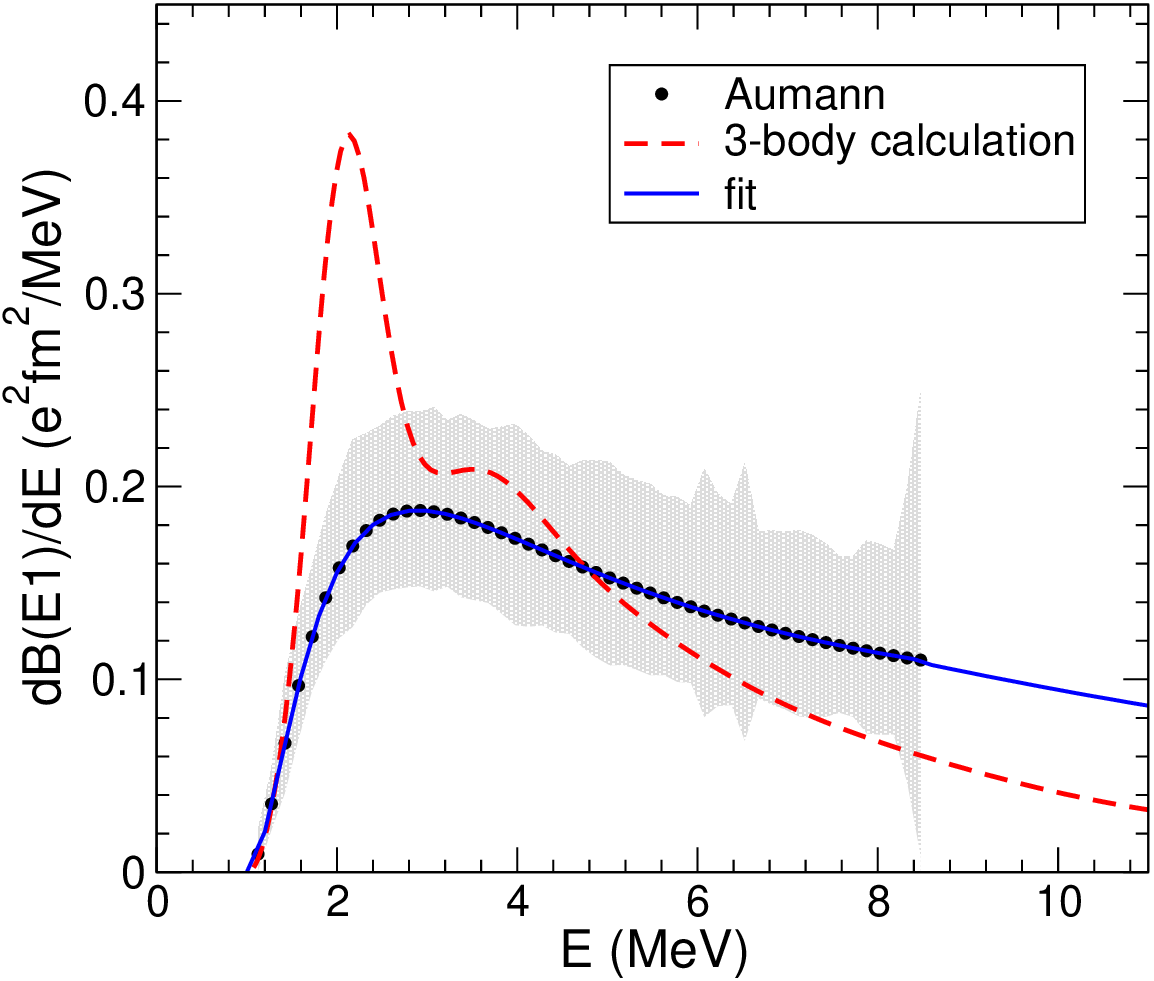}\end{center}%
\caption{\label{fig1}
Electric dipole line strength $dB(E1)/dE$ in units $e^2\,{\rm fm}^2/$MeV for $^6$He. The 
solid line is a fit to the experimental distribution of  Aumann {\em et al.} \cite{Aum99} 
(filled circles), extrapolated to 12 MeV. The shadow 
region indicates the experimental uncertainty of the data. The dashed line is the 
calculation of Thompson  {\em et al.} \cite{Tho00}.}
\end{figure}

We first consider the $^6$He isotope. In this case the polarizability 
$\tilde \alpha(^6{\rm He})$ was calculated from two different $B(E1)$ distributions. The 
first one corresponds to the experimental distribution extracted by Aumann {\em et al.} 
\cite{Aum99}
from the  Coulomb breakup of $^6$He on  lead  at 240 MeV/u. These data are represented 
by the dots in Fig.~\ref{fig1}. The second $B(E1)$ distribution corresponds to a 
theoretical calculation, and was obtained in  \cite{Tho00} 
by  evaluating the matrix element of the dipole operator between the ground state and 
the $1^-$ continuum states. These states where obtained by solution of the Schr\"odinger 
equation, assuming a three-body model for the  $^6$He nucleus, with pairwise 
neutron-neutron and neutron-$\alpha$ interactions, plus an  effective three-body force. 
The $B(E1)$ obtained in this calculation is represented in Fig.~\ref{fig1} by the 
dashed line. 
In spite of the discrepancy between the theoretical  and the 
experimental $B(E1)$ distributions, 
the deduced polarizability $\tilde\alpha_{\rm pol}$, as obtained from Eq.~(\ref{03}),
is similar in both cases: $\tilde\alpha_{\rm pol, the} = 24.2\,{\rm fm}^3$
versus  $\tilde\alpha_{\rm pol, exp} = 21.1\,{\rm fm}^3$. It should be noted 
that the experimental data were extrapolated and integrated up to $E_T = 12.3$
MeV, the threshold value for the decay into two tritons. We do not have,
however, experimental data for the $B(E1)$ beyond this threshold. Therefore,
for the final result
we take the average and add the polarizability of $^4$He [calculated 
in Eq.~(\ref{14})] to partially account for other decay channels, and obtain
\begin{equation}
\tilde \alpha_{\rm pol}(^6{\rm He}) = 24.7(5.0)\,{\rm fm}^3 = 4.3(9)\cdot10^{-7}\;
m^{-3} \, , \label{09}
\end{equation}
where $m$ is the electron mass. For the comparison
with other possible determinations of $^6$He polarizabilities,
we additionally present in Table I the static electric dipole and logarithmically modified
polarizabilities. However, they can not be used to determine
$\tilde\alpha_{\rm pol}$ since the $E_T$ is of the order of the electron mass $m$.
\begin{table}[hbt]
\label{pol}
\begin{tabular}{llw{1.7}w{1.7}w{1.7}}
\hline\hline                   & Ref.
                               & \centt{$\alpha_{\rm E}$[fm$^3$]}    
                               & \centt{$\alpha_{\rm Elog}$[fm$^3$]} 
                               & \centt{$\tilde\alpha_{\rm pol}$[fm$^3$]}\\ \hline
$^3$He                        & &   0.153(15) & 0.615(62) & 3.56(36)\\
&\cite{rinker}                 &   0.130(13)  &           &         \\
&\cite{leid}                   &   0.145      &           &         \\
&\cite{goec}                   &   0.250(40)  &           &         \\ \hline
$^4$He                        & &   0.076(8)  & 0.365(37) & 2.07(20)\\
&\cite{leid}                   &   0.076      &           &         \\
&\cite{gazit}                  &   0.0655     &           &         \\
&\cite{friar_m}                &   0.072(4)   &           &         \\ \hline
$^6$He                        & &   1.99(40)  & 4.78(96)  & 24.7(5.0)\\
\hline\hline\\
\end{tabular}
\caption{The electric dipole polarizability $\alpha_{\rm E}$, 
         the logarithmically modified polarizability $\alpha_{\rm Elog}$, and
         the weighted polarizability $\tilde\alpha_{\rm pol}$ for helium isotopes. }
\end{table}
The resulting contribution to the  frequency  
of, for example, the $2^3S_1 - 3^3P_2$ transition in $^6$He is
\begin{eqnarray}
E_{\rm pol} &=& -m\,\alpha^4\,\Bigl\langle\delta^3(r_1) +\delta^3(r_2) 
\Bigr\rangle_{2^3S_1 - 3^3P_2}\;(m^3\,\tilde\alpha_{\rm pol}) = h\,\nu_{\rm
  pol} \,\, . \label{10}\\
\nu_{\rm pol} &=& 0.015(3) \,{\rm MHz} \, . \label{11}
\end{eqnarray}
For comparison, the finite nuclear size correction to the same transition is \cite{Wan04}:
\begin{eqnarray}
E_{\rm fs} &=&\frac{2\,\pi}{3}\,Z\,\alpha^4\,m^3\,\langle r_{\rm ch}^2\rangle\,
\Bigl\langle\delta^3(r_1) +\delta^3(r_2)\Bigr\rangle_{2^3S_1 - 3^3P_2} =
h\,\nu_{\rm fs} \,\, . \label{12}\\
\nu_{\rm fs} &=&  -1.008\,\langle r_{\rm ch}^2\rangle \frac{\rm MHz}{{\rm
    fm}^2} = -4.253\,{\rm MHz}  \,\, . \label{13}
\end{eqnarray}
The relative magnitude of the nuclear polarizability to the nuclear finite
size for $^6$He is about $0.35$~\%, so it alters the charge radius
determination at this precision level. However, the uncertainty of the experimental result
of Wang {\em et al.} \cite{Wan04} for the isotope shift between $^6$He and $^4$He,
$\nu_{\rm iso} = 43\,194\,772(56)$ kHz,
is about four times larger than $\nu_{\rm pol}$, and therefore the nuclear polarizability correction
at this precision level is not very significant.

We proceed  now to evaluate the  $^4$He nuclear polarizability. This has been 
obtained from the total photoabsorption cross section  measured by 
Arkatov {\em et al.} \cite{Ark78,Ark80,Fuh95}.
\begin{figure}[htb]%
\begin{center}\includegraphics[width=0.7\linewidth]{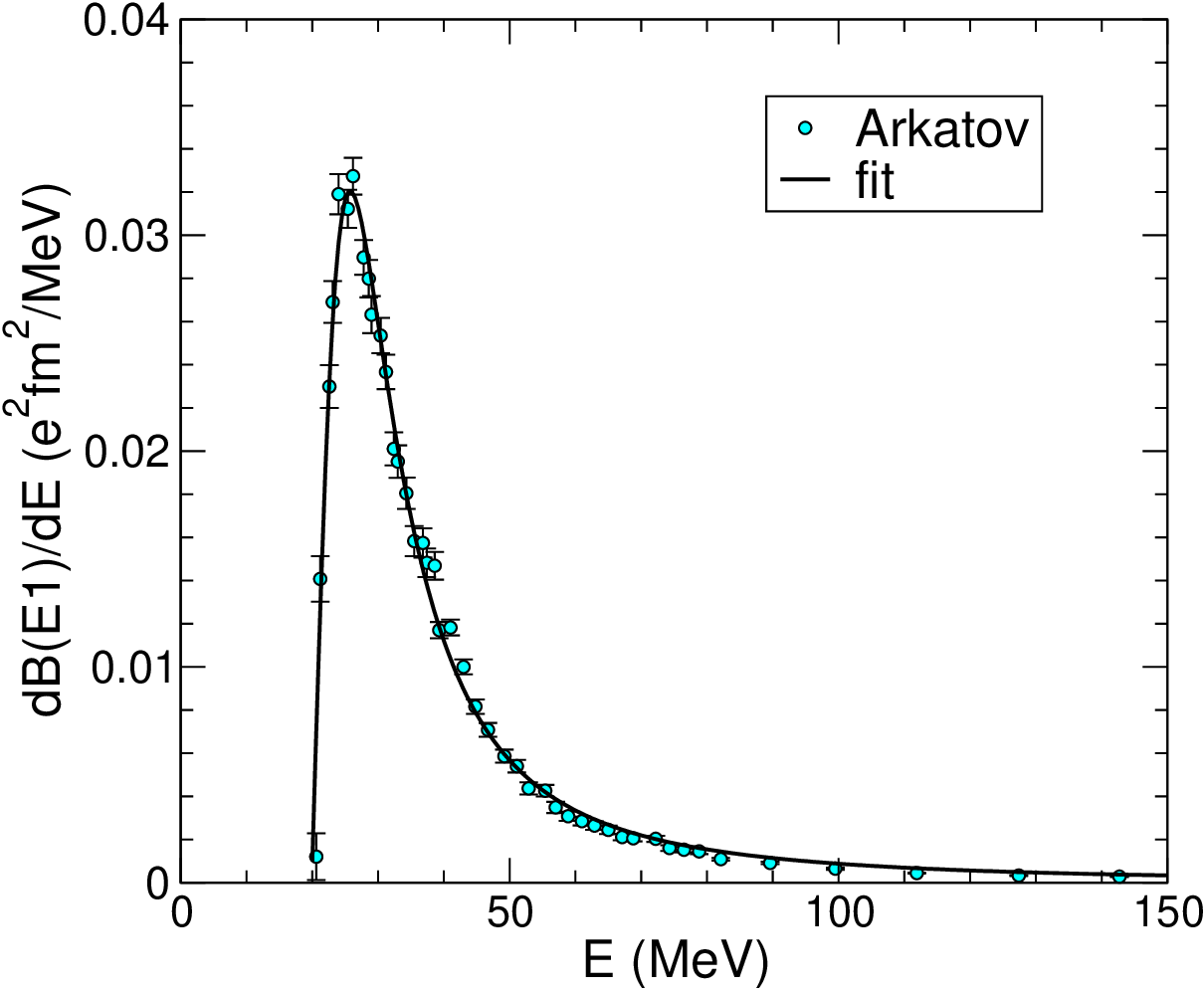}\end{center}%
\caption{\label{fig2}
Electric dipole line strength $dB(E1)/dE$ in units $e^2\,{\rm fm}^2/$MeV for $^4$He. The circles
correspond to the data of Arkatov {\em et al.} \cite{Ark78,Ark80,Fuh95}. The solid line is 
a fit to the data. }
\end{figure}
Using  Eq.~(\ref{05}),  
we extracted from these data the $B(E1)$  distribution shown
in Fig.~\ref{fig2} (filled circles). Then, applying  Eq.~(\ref{06}), one obtains the 
weighted polarizability of $^4$He:
\begin{equation}
\tilde\alpha ({^4}{\rm He}) = 2.07(20)\,{\rm fm}^3 = 3.6(4)\, 10^{-8}\,m^{-3} \, . \label{14}
\end{equation}
It should be noted that at $100$ MeV the dipole approximation in Eq.~(\ref{01}) may not be
correct, since the corresponding photon wavelength is of order of the nuclear size.
Therefore we introduce $10\%$ uncertainty to account for this approximate
treatment. Results for the static polarizability obtained in this work, along with
those obtained in other works, are presented in Table I. 
Our static polarizability agrees with that of
Friar \cite{friar_m}, which was based on earlier experimental data
for the photoproduction cross section. It agrees also with theoretical
calculations by Leidemann \cite{leid}, but slightly disagrees with the most
recent calculations of Gazit {\em et al} \cite{gazit}.
 
The obtained weighted polarizability gives a relatively small effect for transitions in neutral helium.
The ratio of the polarizability  to the finite size correction is 
$4.6\cdot 10^{-4}$. At present, there are no measurements for the charge
radii at this precision level. However, from the 
planned high precision measurement of the $1S-2S$ transition in $^4$He$^+$ \cite{hansch},
one  can in principle obtain the charge radius 
of the $\alpha$ particle from the knowledge of the  nuclear polarizability. The 
polarizability correction to this transition amounts to
\begin{equation}
\nu_{\rm pol}(1S-2S, {\rm ^4He}^+) = 28(3)\, {\rm kHz} \label{15}
\end{equation}
and is smaller than the uncertainty of about $400$ kHz 
in current theoretical predictions, which are due to $\alpha^2\,(Z\,\alpha)^6$
higher order two-loop electron self-energy corrections \cite{yerokhin}.

For the $^3$He atom, we separately calculate the nuclear polarizability corrections 
coming from the two- and three-body photodisintegration processes. 
There are many experimental results for the photoabsorption cross section as well as 
theoretical calculations. They agree fairly well for the two-body dissociation,
but they significantly differ  for the case of three-body disintegration
at high excitation energies ($E_\gamma > 20$ MeV). Since these experimental results have large
uncertainties we prefer to rely on theoretical calculations which agree with
each other very well. In this work, we will use the calculations of 
Deltuva {\em et al.} \cite{Del05} and Golak {\em et al.}
\cite{Gol05}.  The former uses the realistic CD Bonn NN potential, 
supplemented with a three-body force to 
account for the $\Delta$ excitation, and a full treatment of the  Coulomb potential. The calculation 
of Golak {\em et al.}  is based 
on the AV18 NN interaction in combination with the UrbanaIX three-nucleon force, and considers 
Coulomb effects just
for the ground state.  The  $B(E1)$ distributions, extracted from these theoretical  
photodisintegration cross sections, are displayed in Figs.~3 and  4, respectively,  
as a function of the excitation energy. 
\begin{figure}[htb]%
\begin{center}\includegraphics[width=0.7\linewidth]{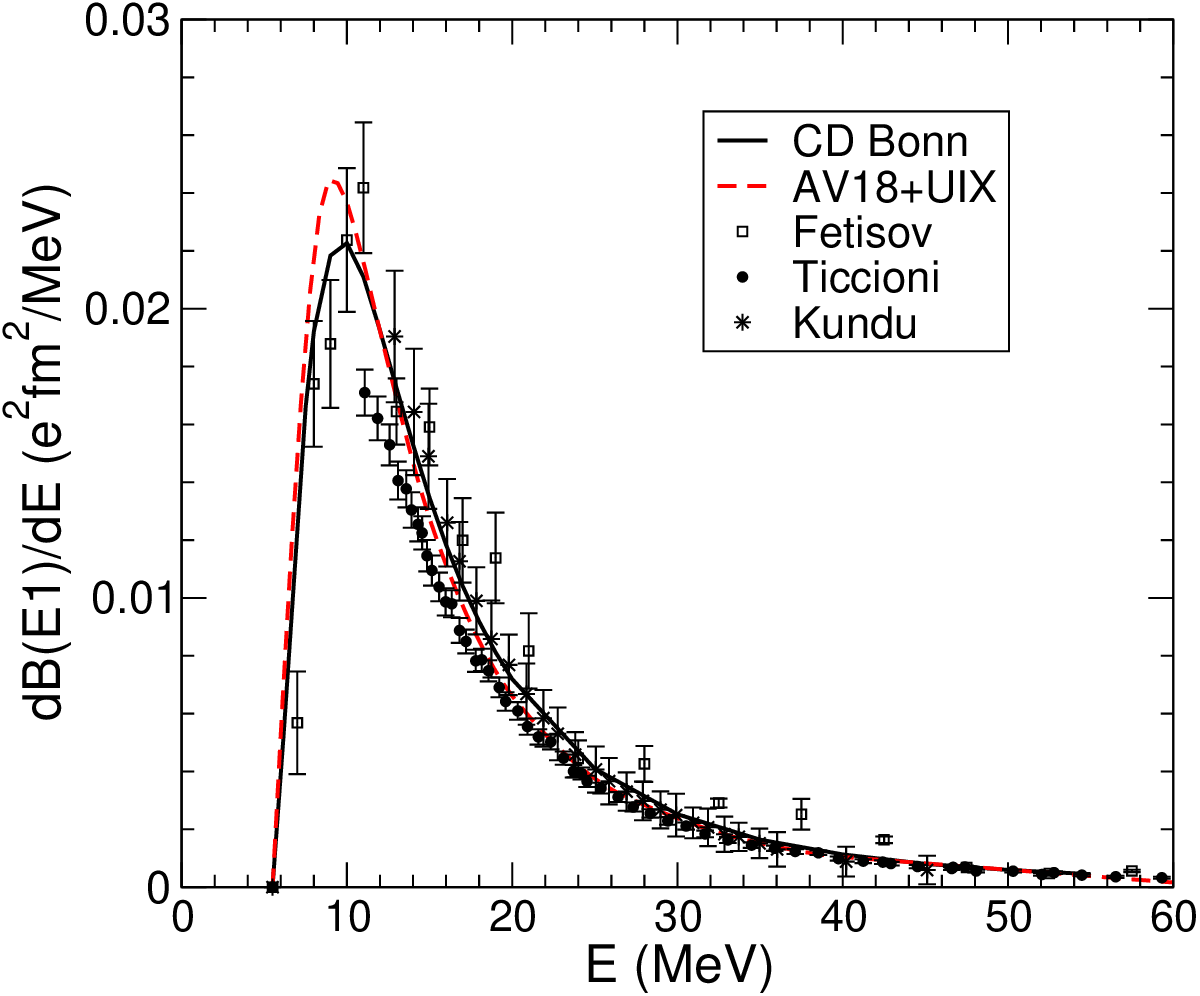}\end{center}%
\caption{\label{fig3}
Electric dipole line strength $dB(E1)/dE$ in units $e^2\,{\rm fm}^2/$MeV for 
two-body disintegration of $^3$He. The data are from Fetisov {\em et al.} \cite{Fet65}, 
Ticcioni {\em et al.} \cite{Tic73} and Kundu {\em et al.} \cite{Kun71}. 
The solid and dashed lines are 
the calculations by Deltuva {\em et al.} \cite{Del05} and Golak {\em et al.}
\cite{Gol05}, respectively.}
\end{figure}
\begin{figure}[htb]%
\begin{center}\includegraphics[width=0.7\linewidth]{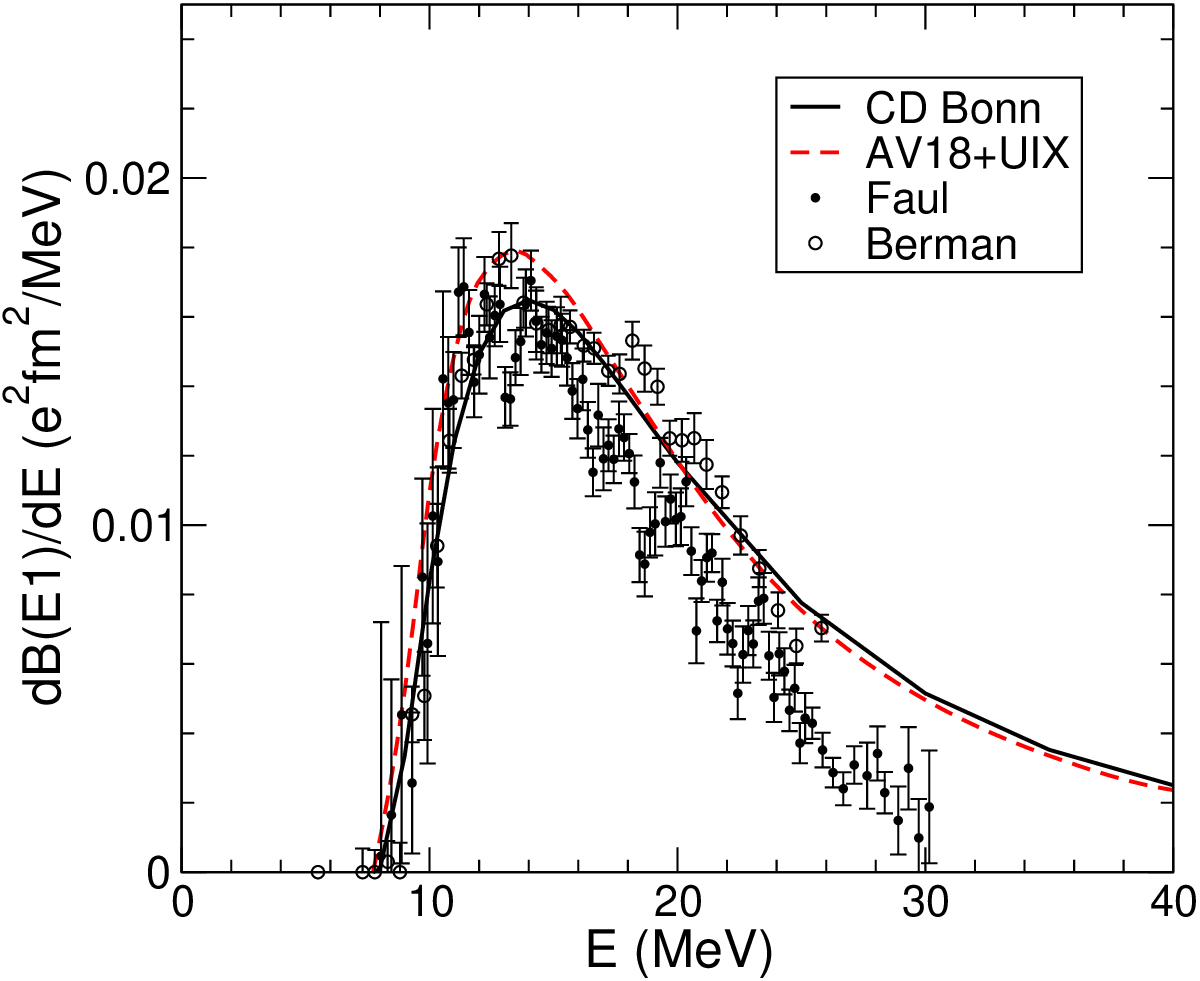}\end{center}%
\caption{\label{fig4}
Electric dipole line strength $dB(E1)/dE$ in units $e^2\,{\rm fm}^2/$MeV for 
three-body disintegration of $^3$He. The data are from Faul {\em et al.} \cite{Fau81} and
Berman {\em et al.} \cite{Ber74}. The solid and dashed lines are the calculations by 
 Deltuva {\em et al.} \cite{Del05} and Golak {\em et al.} \cite{Gol05}, respectively.}
\end{figure}

Using these theoretical $B(E1)$ distributions, the result for the weighted
polarizability due to both two- and three-body
disintegration is
\begin{equation}
\tilde\alpha ({\rm ^3He})= 3.56(36)\,{\rm fm}^3 = 6.2(6)\, 10^{-8}\,m^{-3} , \label{16}
\end{equation} 
and the related contribution to the frequency of the $1S-2S$ transition in He$^+$ is 
\begin{equation}
\nu_{\rm pol}(1S-2S,{\rm ^3He}^+) = 48(5)\, {\rm kHz} . \label{17}
\end{equation}
Hence, the magnitude of the nuclear polarizability correction for $^3$He is almost
twice as large as for $^4$He and can be significant for the absolute charge radius
determination from the $1S-2S$ measurement in hydrogen-like helium.
The corresponding contribution to the $^4$He-$^3$He isotope shift in the $2^3S_1-2^3P_J$
transition of $-1$ kHz is at present much smaller than the experimental 
precision of about $30$ kHz \cite{3he}. 

Our result for the static polarizability of $^3$He is presented in Table I. 
It is in agreement with the first calculation by Rinker \cite{rinker}
which was based on measured that time photoabsorption cross section,
it is also in agreement with calculations of Leidemann
\cite{leid}, but is in disagreement with the cross section mesurement 
for the elastic scattering of $^3$He nuclei of $^{208}$Pb below the Coulomb
barrier \cite{goec}. 

In summary, we have obtained the nuclear polarizability correction in 
helium isotopes. In most cases, we find that
the correction to the energy levels is smaller than current experimental
precision, but could affect the determination of the charge
radius when more accurate measurements become available.
Together with possible high precision measurements in muonic atoms, 
it will allow for an improved test of QED and a more accurate determination of the
fundamental constants. 

\section*{Acknowledgments}
We wish to acknowledge fruitful discussions with K. Rusek, A. Deltuva and F. Kottmann. 
We are grateful to T. Aumann and R. Skibi\'nski for providing us the He data 
in tabular form.


\begin{thebibliography}{99}

\bibitem{bj} J. Bernab\'{e}u and C. Jarlskog, Nucl. Phys. {\bf B47}, 205 (1974).

\bibitem{friar_m} J.L. Friar, Phys. Rev. C {\bf 16}, 1540 (1977).

\bibitem{huber} A. Huber, {\em et al}., Phys. Rev. Lett. {\bf 80}, 468 (1998).

\bibitem{11li} R. S\'anchez, {\em et al}., Phys. Rev. Lett. {\bf 96}, 033002 (2006).

\bibitem{ros} W. Leidemann and R. Rosenfelder, Phys. Rev. C {\bf  51}, 427 (1995).

\bibitem{friar} J.L. Friar and G.L. Payne, Phys. Rev. C {\bf 56}, 619 (1997). 

\bibitem{mp} M. Puchalski, A. Moro and K. Pachucki, Phys. Rev. Lett. {\bf 97}, 133001 (2006).

\bibitem{weitz} K. Pachucki, D. Leibfried, and T.W. H\"ansch, Phys. Rev. A {\bf 48}, R1 (1993);
               K. Pachucki, M. Weitz, and T.W. H\"ansch, Phys. Rev. A {\bf 49}, 2255 (1994).

\bibitem{audi}  G. Audi, A.H. Wapstra, and C. Thibault, Nucl. Phys. {\bf A729}, 337 (2003).

\bibitem{Aum99} T. Aumann {\em et al.}, Phys. Rev. C {\bf 59}, 1252 (1999).

\bibitem{Tho00}  I.J. Thompson, B.V. Danilin, V.D. Efros, J.S. Vaagen,
             J.M. Bang and M.V. Zhukov, Phys. Rev. C {\bf 61}, 024318 (2000).

\bibitem{rinker} G.A. Rinker, Phys. Rev. A {\bf 14}, 18 (1976).

\bibitem{leid} W. Leidemann, Few-Body Syst. Suppl. {\bf 14}, 313 (2003).

\bibitem{goec} F. Goeckner, L.O. Lamm, and L.D. Knutson, Phys. Rev. C {\bf 43}, 66 (1991).

\bibitem{gazit} D. Gazit, N. Barnea, S. Bacca, W. Leidemann, and G. Orlandini,
                Phys. Rev. C {\bf 74}, 061001(R) (2006);
                D. Gazit, S. Bacca, N. Barnea, W. Leidemann, and G. Orlandini,
                 Phys. Rev. Lett. {\bf 96}, 112301 (2006).

\bibitem{Wan04} L.-B. Wang {\em et al.}, Phys. Rev. Lett. {\bf 93}, 142501 (2004).

\bibitem{Ark78}  J.M. Arkatov, P.I. Vatset, V.I. Voloshchuk, 
                 V.N. Gurev, A.F. Khodyachikh, Ukr. Fiz. Zh. {\bf 23}, 1818 (1978).

\bibitem{Ark80} Yu.M. Arkatov, P.I. Vatset, V.I. Voloshchuk, V.A. Zolenko,
                I.M. Prokhorets, Yad. Fiz. {\bf 31}, 1400 (1980); 
                [Sov. J. Nucl. Phys. {\bf 31}, 726 (1980)].

\bibitem{Fuh95} K. Fuhrberg {\em et al.}, Nucl. Phys. A {\bf 591}, 1 (1995).

\bibitem{hansch}  G. Gohle {\em et al.}, Nature {\bf 436}, 234 (2005).

\bibitem{yerokhin} V.A. Yerokhin, P. Indelicato, and V.M. Shabaev,
                   Phys. Rev. A {\bf 71}, 040101(R) (2005).

\bibitem{Del05} A. Deltuva, A.C. Fonseca, and P.U. Sauer, Phys.~Rev.~C {\bf 71},  054005  (2005); 
                  {\em ibid} {\bf 72},  054004  (2005).

\bibitem{Gol05}  J. Golak, R. Skibi\'nski, H. Wita\l a, W. Gl\"ockle, A. Nogga, H. Kamada, 
                 Phys. Rep. {\bf 415}, 89 (2005). 

\bibitem{Fet65}  V.N. Fetisov, A.N. Gorbunov, and  A.T.Varfolomeev,
                 Nucl. Phys. {\bf 71}, 305 (1965).

\bibitem{Tic73}  G. Ticcioni, S.N. Gardiner, J.L. Matthews, and R.O. Owens,
                 Phys. Lett. {\bf 46B}, 369 (1973).

\bibitem{Kun71}  S.K. Kundu, Y.M. Shin, G.D. Wait, Nucl. Phys. A {\bf 171}, 384 (1971).

\bibitem{Fau81}  D.D. Faul, B.L. Berman, P. Meyer, and D.L. Olson, Phys. Rev. C {\bf 24}, 849 (1981).

\bibitem{Ber74}  B.L. Berman, S.C. Fultz, P.F. Yergin, Phys. Rev. C {\bf 10}, 2221 (1974).

\bibitem{3he} D.C. Morton, Q. Wu, and G.W.F. Drake, Phys. Rev. A {\bf 73}, 034502 (2006).


\end{thebibliography}
\end{document}